# Possible quantum gravity effects in a charged Bose condensate under variable e.m. field

Giovanni Modanese

*I.N.F.N. - Gruppo Collegato di Trento*

*Dipartimento di Fisica dell'Università*

*I-38050 Povo (TN) - Italy*

E-mail: giovanni.modanese@unibz.it

*and*

John Schnurer

*Physics Engineering*

*P.O. Box CN 446, Yellow Springs, Ohio 45387-0466 - U.S.A.*

## Abstract

In the weak field approximation to quantum gravity, a "local" positive cosmological term $\mu^2(x)$ corresponds to a local negative squared mass term in the Lagrangian and may thus induce instability and local pinning of the gravitational field. Such a term can be produced by the coupling to an external Bose condensate. In the functional integral, the local pinning acts as a constraint on the field configurations. We discuss this model in detail and apply it to a phenomenological analysis of recent experimental results.

## Resume

Dans l'approximation a champ faible de la gravite quantique, a un terme "local" cosmologique positif $\mu^2(x)$ corresponds une masse locale imaginaire dans le Lagrangian et il peut donc engendrer instabilité et piégeage locale du champ gravitationnel. Ce terme peut être produit par le couplage avec un condensat de Bose extérieur. Dans l'intégrale fonctionnelle, le piégeage local agit comme une contrainte sur les configurations du champ. Nous discutons ce model en détail et nous l'appliquons a une analyse phenomenologique des résultats expérimentaux récents.

The behavior of a Bose condensate - or more specifically of the Bose condensate of Cooper pairs within superconducting materials - in an external gravitational field has been the subject of some study in the past [1]. The presence in a superconductor of currents flowing without any measurable resistance suggests that a superconductor could be used as a sensitive detection system, in particular for gravitational fields. The possible back-reaction of induced supercurrents on the gravitational field itself has been studied too, in analogy with the familiar treatment of the Meissner effect. As one would expect, it turns out that the "gravitational Meissner effect" is extremely weak: for instance, it has been computed that in a neutron star with a density of the order of $10^{14}$ $g/cm^3$, the London penetration depth is ca. 12 $km$ [2].

Such a penetration depth shows that the coupling of supercurrents to the classical gravitational field is extremely weak. The reason, of course, stems from the smallness of the coupling between gravity and the energy-momentum tensor of matter $T_{\mu\nu}$. One might wonder whether in a quantum theory of gravity - or at least in an approximation of the theory for weak fields - the Bose condensate of the Cooper pairs, due to its macroscopic quantum character, can play a more subtle role than as a simple contribution to the energy-momentum tensor.

In other words, we wonder if the macroscopic quantum coherence of the condensate can be taken into account at a fundamental level when computing the interaction between the superconductor and the external gravitational field. Results might then differ from those obtained from the gravitational coupling of "regular", incoherent matter. A similar approach has been earlier proposed by Weber in his studies of gravitational waves and neutrino fluxes detection [3].

In a quantum-field representation the condensate is described by a field with non-vanishing vacuum expectation value $\phi_0$, possibly depending on the spacetime coordinate $x$. It is interesting to insert the action of this field, suitably "covariantized", into the functional integral of gravity, expand the metric tensor $g_{\mu\nu}$ in weak field approximation, and check if the only effect of $\phi_0(x)$ is to produce gravity/condensate interaction vertices proportional to powers of $\kappa = \sqrt{(16\pi G)}/c^2$. One finds that in general this is not the case; the quadratic part of the gravitational action is modified as well, by receiving a negative definite contribution. It can thus be expected that the condensate induces localized instabilities of the gravitational field, in a sense which we shall describe precisely in Section 2.

The present paper is based on the letter [4] and originates in part from previous theoretical work [5] and in part from recent experimental results ([6]; see Sections 3, 4) which show the possibility of an anomalous interaction, in special conditions, between the gravitational field and a superconductor. In Section 1 the theoretical model is introduced. In Section 1.1 we recall the role of a global cosmological term in the gravitational action. In Section 1.2 we show how a "local" positive cosmological term appears in the gravitational action due to the minimal coupling to a scalar field $\Phi(x)$ with $x$-dependent vacuum expectation value $\phi_0(x)$, representing a Bose condensate with variable density. We define the "critical regions" of spacetime as those in which the overall sign of the cosmological term is positive. In Section 2 we discuss the gravitational instability induced by the condensate in the critical regions. We note that the "kinetic term" $R(x)$ in the gravitational lagrangian admits zero modes, and discuss qualitatively the effect of the instabilities in the Euclidean functional integral for the gravitational static potential energy [5]. In Section 3 we summarize the experimental findings reported in [6], trying to focus on the essential elements of an experimental set-up, which is quite complex. We add some remarks on the possible theoretical interpretations of these observations, according to the model with "anomalous" coupling between $h(x)$ and $\phi_0(x)$ introduced in the previous Sections. In Section 4 we describe our own experimental results and their interpretation. Finally, in Section 5 we discuss some issues of elementary character - but important from the practical point of view - concerning the overall energetic balance of the shielding process.

Throughout this paper CGS units are employed.

# 1 Bose condensate as "local" cosmological term in perturbative quantum gravity.

In this Section we show that a scalar field with a $x$-dependent vacuum expectation value, coupled to the gravitational field, gives a positive $x$-dependent contribution to the cosmological term. Our argument follows that given in [7].

### 1.1 Global cosmological term.

Let us consider the action of the gravitational field $g_{\mu\nu}(x)$ in its usual form [8]:

$$S_g = \int d^4x \sqrt{g(x)} c^4 [(\Lambda/8\pi G)-(R(x)/8\pi G)] \qquad (1)$$

where $-c^4 R(x)/8\pi G$ is the Einstein term and $c^4 \Lambda/8\pi G$ is the cosmological term which generally can be present.

It is known that the coupling of the gravitational field with another field is formally obtained by "covariantizing" the action of the latter; this means that the contractions of the Lorentz indices of the field must be effected through the metric $g_{\mu\nu}(x)$ or its inverse $g^{\mu\nu}(x)$ and that the ordinary derivatives are transformed into covariant derivatives by inserting the connection field. Moreover, the Minkowskian volume element $d^4x$ is replaced by $d^4x\sqrt{g(x)}$, where $g(x)$ is the determinant of the metric. The insertion of the factor $\sqrt{g(x)}$ into the volume element has the effect that any additive constant in the Lagrangian contributes to the cosmological term $c^4$

$\Lambda/8\pi G$. For instance, let us consider a Bose condensate described by a scalar field $\Phi(x)=\phi_0 + \phi(x)$, where $\phi_0$ is the vacuum expectation value and $m_\Phi c^2|\phi_0|^2$ represents the particles density of the ground state in the non-relativistic limit (compare eq.s (8)-(10)). The action of this field in the presence of gravity is

$$S_\Phi = \int d^4x \sqrt{g(x)} \{c^2 [\partial_\mu \phi(x)]^* [\partial_\nu \phi(x)] g^{\mu\nu}(x) + m_\Phi^2 c^4|\phi(x)|^2 + m_\Phi^2 c^4 [\phi_0^* \phi(x) + \phi^*(x) \phi_0] + m_\Phi^2 c^4 |\phi_0|^2 \}. \qquad (2)$$

Considering the total action ($S_g+S_\Phi$), it is easy to check that the "intrinsic" gravitational cosmological constant $\Lambda$ receives a contribution $(1/2)m_\Phi^2 c^4|\phi_0|^2 8\pi G$.

Astronomical observations impose a very low limit on the total cosmological term present in the action of the gravitational field. The presently accepted limit is on the order of $|\Lambda|G\hbar/c^3 < 10^{-120}$, which means approximately for $\Lambda$ itself, that $|\Lambda| < 10^{-54}$ $cm^{-2}$. (See [9] and references therein.) This absence of curvature in the large-scale universe rises a paradox, called "the cosmological constant problem" [10]. In fact, the Higgs fields of the standard model as well as the zero-point fluctuations of any quantum field, including the gravitational field itself, generate huge contributions to the cosmological term - which, somehow, appear to be "relaxed" to zero at macroscopic distances. In order to explain how this can occur, several models have been proposed [11]. No definitive and universally accepted solution of the paradox seems to be at hand, if only because that would require a complete non-perturbative treatment of gravity - which for the moment, at least, is not feasible [12].

A model in which the large scale vanishing of the effective cosmological constant has been reproduced through numerical simulations is the Euclidean quantum gravity on the Regge lattice [13]. From this model a property emerges, that could turn out to be more general than the model itself: if we keep the fundamental length $L_{Planck}=\sqrt{(G\hbar/c^3)}$ in the theory, the vanishing of the effective cosmological constant $|\Lambda|$, depending on the momentum scale $p$, follows a law of the form $|\Lambda|(p) \sim c^3(G\hbar)^{-1}(L_{Planck} p/\hbar)^\gamma$, where $\gamma$ is a critical exponent [14,15]. We find no reason to exclude that this behavior of the effective cosmological constant may be observed in certain circumstances (see [7]). Furthermore, the model predicts that in the large distance limit, $\Lambda$ goes to zero while keeping negative sign. Also, this property probably has a more general character, since the weak field approximation for the gravitational field is "stable" in the presence of an infinitesimal cosmological term with negative sign; conversely, it becomes unstable in the presence of a positive cosmological term (see [16] and Section 2).

## 1.2 Local cosmological term.

Summarizing, there appears to exist, independent of any model, a dynamical mechanism that "relaxes to zero" any contribution to the cosmological term. This makes the gravitational field insensitive to any constant term in the action of other fields coupled to it. Nevertheless, let us go back to the previously mentioned example of a Bose condensate described by a scalar field $\Phi(x)=\phi_0 + \phi(x)$. If the vacuum expectation value $\phi_0$ is not constant but depends on the spacetime coordinate $x$, then a positive "local" cosmological term appears in the gravitational action $S_g$, and this can have important consequences. Let us suppose that $\phi_0(x)$ is fixed by external factors. Now, let us decompose the gravitational field $g_{\mu\nu}(x)$ as usual in the weak field approximation, that is, $g_{\mu\nu}(x)=\delta_{\mu\nu}+ h_{\mu\nu}(x)$. The total action of the system takes the form

$$S = \int d^4x \sqrt{g(x)} \, c^4[(\Lambda/8\pi G)+(\mu^2(x)/2) - (R(x)/8\pi G)] + S_{h,\phi_0} + S_\phi, \qquad (3)$$

where

$$\mu^2(x) = c^2 \, [\partial_\mu \phi_0^*(x)] \, [\partial^\mu \phi_0(x)] + m_\Phi^2 \, c^4 \, |\phi_0(x)|^2 \qquad (4)$$

$$S_{h,\phi_0} = \int d^4x \sqrt{g(x)} \, c^2 \, \partial^\mu \phi_0^*(x) \, \partial^\nu \phi_0(x) \, h_{\mu\nu}(x) \qquad (5)$$

$$S_\phi = \int d^4x \sqrt{g(x)} \, \{m_\Phi^2 \, c^4 |\phi(x)|^2 + m_\Phi^2 \, c^4 \, [ \, \phi_0^*(x) \, \phi(x) + \phi_0(x) \, \phi^*(x) \, ] +$$

$$+ c^2 \, [ \, \partial_\mu \phi^*(x) \, \partial_\nu \phi(x) + \partial_\mu \phi_0^*(x) \, \partial_\nu \phi(x) + \partial_\mu \phi^*(x) \, \partial_\nu \phi_0(x) \, ] \, g^{\mu\nu}(x)\} \qquad (6)$$

In the action $S$, the terms $S_{h,\phi_0}$ and $S_\phi$ represent effects of secondary importance. The term $S_{h,\phi_0}$ describes a process in which gravitons are produced by the "source" $\phi_0(x)$. The term $S_\phi$ contains the free action of the field $\phi(x)$, describing the excitations of the condensate and several vertices in which the graviton field $h_{\mu\nu}(x)$ and $\phi(x)$ interact between themselves and possibly with the source. None of these interactions is of special interest here; they are generally very weak, due to the smallness of the gravitational coupling $8\pi G/c^4$. The relevant point (eq.s (3), (4)) is that the purely gravitational cosmological term $[(c^4\Lambda)/(8\pi G)]$ receives a local positive contribution $(1/2)\mu^2(x)$ that depends on the external source $\phi_0(x)$.

We shall use "critical regions" to designate those regions of spacetime where the following condition is satisfied:

$$(c^4\Lambda/8\pi G) + (\mu^2(x)/2) > 0. \qquad (7)$$

As we shall see in Section 2, in these regions the gravitational lagrangian is unstable and the field tends to be "pinned" at some fixed values.

It is important to give a numerical estimate of the magnitude order of $\mu^2(x)$ in the case of a superconductor. To this end we recall that the Hamiltonian of a scalar field $\Phi$ of mass $m_\Phi$ is given by

$$H = (1/2) \int d^3x \, \{|\partial \Phi(x)/\partial t|^2 + c^2 \sum_{i=1,2,3}|\partial \Phi(x)/\partial x^i|^2 + m_\Phi^2 \, c^4 \, |\Phi(x)|^2\}. \qquad (8)$$

In our case $\Phi$ describes a system with a condensate and its vacuum expectation value is $\langle 0|\Phi(x)|0\rangle = \phi_0(x)$. Then in the Euclidean theory $\mu^2$ is positive definite and we have

$$\langle 0|H|0\rangle = (1/2) \int d^3x \, \mu^2(x). \qquad (9)$$

Thus $\mu^2$ is connected to the coherent vacuum energy density. A prudent estimate gives, for a superconductor

$$\mu^2 > 10 \ erg/cm^3. \qquad (10)$$

As we saw in Section 1.1, a typical upper limit on the intrinsic cosmological constant observed at astronomical scale is $|\Lambda| G\hbar/c^3 < 10^{-120}$, which means $|c^4\Lambda|/8\pi G < 2\cdot 10^{-7} \ erg/cm^3$. This small value, compared with the above estimate for $\mu^2(x)$, supports our assumption that the total cosmological term can assume positive values in the superconductor, and that the criticality condition can be satisfied in some regions. But in fact the positive contribution of the condensate is so large that one could expect the formation of gravitational instabilities in any superconductor or superfluid, subjected to external e.m. fields or not - a conclusion in contrast with the observations.

According to our discussion in Section 1.1, we may then hypothesize that the value of $c^4 \Lambda/8\pi G$ at small scale is larger than that observed at astronomical scale, and negative in sign, such that it would represent a "threshold" of the order of $\sim 10 \ erg/cm^3$ for anomalous gravitational couplings. As we noted in Section 1.1, a negative intrinsic cosmological constant is allowed in models of quantum gravity containing a fundamental length. Given the small magnitude cited above, this threshold would not influence any other known physical process.

## 2  Gravitational instability and field "pinning" in the critical regions.

In this Section we discuss the gravitational instability induced by the condensate in the critical regions (7). We note that the "kinetic term" $R(x)$ in the gravitational lagrangian admits zero modes. Finally, we discuss the effect of the instabilities in the Euclidean functional integral for the gravitational static potential energy [5].

Let us first consider the Einstein action with cosmological term (1) in the weak field approximation. We recall [16] that after the addition of a harmonic gauge-fixing, the quadratic part of the action takes the form

$$S^{(2)}_g = \int d^4x \ h_{\mu\nu}(x) V^{\mu\nu\alpha\beta} (-\partial^2 - \Lambda) h_{\alpha\beta}(x), \qquad (11)$$

where $V^{\mu\nu\alpha\beta} = \delta^{\mu\alpha}\delta^{\nu\beta} + \delta^{\mu\beta}\delta^{\nu\alpha} - \delta^{\mu\nu}\delta^{\alpha\beta}$.

It is seen from (11) that in this approximation a cosmological term amounts to a mass term for the graviton, positive if $\Lambda < 0$ and negative (with consequent instability of the theory) if $\Lambda > 0$ [17]. Since in the presence of a condensate the sign of the total cosmological term depends on the coordinate $x$ (eq.s (3)-(4)), we expect some "local" instabilities to form in the critical regions (7) in that case.

Such local instabilities were studied in [7], for the case of an Euclidean scalar field $\chi$. Let us summarize very briefly the conclusions. If the squared mass term of the field $\chi$ is negative in some "critical" regions, then inside these regions $\chi$ tends to grow in order to minimize the action. However, this growth is limited by the gradient squared term in the lagrangian. One concludes that the field strength diverges in the critical regions only if the product of the imaginary mass by the size of the critical regions exceeds a certain constant.

In the case of perturbative quantum gravity, the situation is even more favourable to the formation of local instabilities, since the kinetic term $R(x)$ in Einstein action admits zero modes. Among these are not only the field configurations $h_{\mu\nu}(x)$ such that $R(x)=0$ everywhere, but also those which satisfy the Einstein equations

$$R_{\mu\nu}(x) - (1/2)g_{\mu\nu}(x) R(x) = -(8\pi G/c^4) T_{\mu\nu}(x), \qquad (12)$$

with $T_{\mu\nu}$ obeying the condition

$$\int d^4x \sqrt{g(x)} \, g^{\mu\nu}(x) \, T_{\mu\nu}(x) = 0 \qquad (13)$$

(Note that for solutions of (12) one has $R(x)=(8\pi G/c^4) g^{\mu\nu}(x) T_{\mu\nu}(x)$ ).

Condition (13) can also be satisfied by energy-momentum tensors that are not identically zero, but only if they have a balance of negative and positive signs, such that their total integral is zero. Of course, they do not represent any acceptable physical source, but the corresponding solutions of (12) exist nonetheless, and are zero modes of the action. (One can consider, for instance, the static field produced by a "mass dipole".) Clearly, there is no obstacle to the growth of the zero modes in the critical regions [21].

We will admit that the growth of the gravitational field within the critical regions is limited by a positive higher order term in the action, that can be otherwise disregarded. The situation reminds that of a "$\Phi^4 - \Phi^2$" model (see Fig. 1), with the important difference that in the present case the field depends on $x$ and is "pinned" only locally, in the critical regions [17]. We shall denote by $h_{min}$ the "runaway" value of the field $h$, corresponding to the minimum of the double-well potential.

How does this local pinning of the field influences the gravitational interaction of two masses $m_1$ and $m_2$ at rest? We recall the general formula for the static potential in Euclidean quantum gravity [5,14,22]:

$$U(L) = \lim_{\tau \to \infty} (-\hbar/\tau) \log Z^{-1} \int d[g] \exp(-\hbar^{-1} S);$$

$$S = S_g + \sum_{i=1,2} m_i c^2 \int dt \sqrt{\{g^{\mu\nu}[x_i(t)] x_{i\mu}'(t) x_{i\nu}'(t)\}} \qquad (14)$$

where $S_g$ is the gravitational action (1), $Z$ a normalization factor and the integral in $t$ is extended between $(-\tau/2)$ and $(+\tau/2)$. The trajectories $x_1(t)$ and $x_2(t)$ of $m_1$ and $m_2$ are parallel with respect to the metric $g$. $L$ is the distance between the trajectories, corresponding to the spatial distance of the two masses. (Of course, in the case we have in mind, $m_1$ should represent the Earth and $m_2$ a weighed sample, or conversely.)

Eq. (14) means that the interaction energy $U(L)$ of the two masses depends on the correlations between the values of the gravitational field on the "Wilson lines" $x_1(t)$ and $x_2(t)$. This dependence can be verified explicitly in the weak field approximation or through numerical simulations.

We can restate eq. (14) in the presence of the Bose condensate as

$U[L,\mu^2(x)] = \lim_{\tau \to \infty} (-\hbar/\tau) \log Z^{-1} \int d[g] \, d[\phi] \exp(-\hbar^{-1}S)$;

$S = S_{tot}[g,\phi,\phi_0] + \sum_{i=1,2} m_i c^2 \int dt \sqrt{\{g^{\mu\nu}[x_i(t)]x_{i\mu}'(t)x_{i\nu}'(t)\}}$ \hfill (15)

where $S_{tot}[g,\phi,\phi_0]$ is the total action defined in (3). If there are critical regions between the two Wilson lines, the field correlations are modified and in general the field "pinning" reduces $|U|$.

The modification of the static gravitational potential caused by the instability of the field in the critical regions is given by the formula [19]

$U' = U [1 - \gamma ab F(\rho,n)]$, \hfill (16)

where $a$ is the width of a critical region, $b$ its thickness, and $\gamma$ represents the product $\xi h_{min}^2$, being $\xi$ the probability for $h$ to assume the value $h_{min}$. $F$ is an adimensional function of order 1 depending on $\rho = b/a$ and $n=d/a$, where $d$ is the distance of the "shielded" object from the critical region.

This formula can be compared with the experimental data [20]. It turns out that the dependence on $a$, $b$ and $n$ is compatible with the data. The dependence on $\rho$ is very weak and not relevant. The parameter $\gamma = \xi h_{min}^2$, which gives the strength of the effect, cannot be computed theoretically from first principles, because both the runaway strength $h_{min}$ and its probability $\xi$ are unknown. It must in fact be inferred from the experimental data and gives us novel informations about the non-perturbative sector of the theory.

It is therefore impossible, given the present status of the theory, to predict the magnitude order of the effect. The main goal of our theoretical model was that of showing a possible way to elude the "discouraging" weakness of the standard gravitational coupling. The key point was identified in the localized instability induced by the condensate, with subsequent runaway of the field, as explained above.

A more detailed representation of these localized instability was recently given in the context of Lorentzian quantum gravity (which is more generally accepted than the Euclidean formulation) [21]. The metastable modes take here the form of so-called "gravitational dipolar fluctuations", which make the vacuum locally polarizable. Their strength is still defined by the cosmological term, and is not constrained by the small parameter $8\pi G/c^4$ which defines the gravitational coupling to incoherent matter. The probability of their occurrence, however, cannot be computed from the basic elements of the theory.

Coming back to eq. (16), it is interesting to deduce $\gamma$ from the data and work out from this an estimate of $h_{min}$. The width $a$ of the critical regions (coherence domains) in HTC materials lies between $10^{-6}$ and $10^{-4}$ $cm$. The thickness $b$ of the single regions adds up to a total $b$ between $10^{-4}$ and $10^{-2}$ $cm$. Since the observed modification in the gravitational potential is of the order of 1 part in $10^2$, we find for $\gamma$ a value between $10^4$ and $10^8$ $cm^{-2}$. This implies $h_{min} \sim 10^{-29}$ $\xi^{-1/2}$. In this latter formula the strength $h_{min}$ has been written in adimensional units, i.e., $h_{min}$ represents the relative metric modification with respect to flat space. We see that even if the probability $\xi$ of the runaway is very small, the metric modification required by the effect keeps small, too. This confirms that our model is quite reasonable.

# 3  Experimental findings.

A recent experiment [6] has shown an unexpected interaction between the gravitational field and a superconductor subjected to external e.m. fields. In this Section we summarize the reported observations, trying to focus on the essential elements of an experimental setup which is quite complex. We add some remarks on the possible theoretical interpretations of these observations, according to the model with "anomalous" coupling between $h(x)$ and $\phi_0(x)$ introduced in the previous Sections. Attempts at an independent replication of the experiment have been started recently [23].

## 3.1  Summary of results with rotating disks.

The core of the experimental apparatus is a toroidal disk, 27 *cm* in diameter and made of high critical temperature (HTC) superconducting material. The disk is kept at a temperature below 70 *K*; it levitates above three electromagnets and rotates (up to ca. 5000 *rpm*) due to the action of additional lateral magnetic fields. All electromagnets are supplied with variable-frequency AC current.

Within certain frequency ranges one observes a slight decrease in the weight of samples hung above the disk, up to a maximum "shielding" value of ca. 1% (observed when the disk rotates at ca. 5000 *rpm*). A smaller effect, of the order of 0.1% or less, is observed if the disk is only levitating but not rotating. The percentage of weight decrease is the same for samples of different masses and chemical compositions. One can thus describe the effect as a slight diminution of the gravity acceleration $g_E$ above the disk.

The mass of the samples employed in the experiment varied between 10 and 50 *g*. The samples had the form of big pencils hanging in a closed glass tube with the diameter of approximately 3 *cm*. Several different materials were tested: iron, copper, plexiglas, glass, wood, ceramics etc.

This effect does not seem to diminish with increased elevation above the disk: there appears to be a "shielding cylinder" over the disk (compare also [24,25]), that extends upwards for at least 3 meters. The resulting field configuration is clearly non conservative (Fig. 2). An horizontal force at the border of the shielding cylinder has occasionally been observed, but it is far too small to restore the usual "zero circuitation" property of the static field. No weight reduction is observed under the disk. The articles [6] mention several checks performed in order to rule out contributions from e.m. fields, buoyancy and other spurious factors. Thick metal screens were always placed between the samples and the cryostat during the weight measurements. (The same applies to the demonstration experiment described later in this paper.)

The disk has a composite structure: the upper part has been treated with a thermal process which partially melts the grains of the HTC material and makes it a good superconductor, with high $T_c$ (ca. 92 *K*) and high $J_c$ (ca. 7000 $A/cm^2$). On the contrary, the lower part remains more granular and has a lower critical temperature (ca. 65 *K*). The aim of this double structure is to obtain good levitation properties, while also creating a layer in which considerable resistive effects can arise. Both characteristics appear to be necessary for the effect to take place: (1) the disk must be able to support intense super-currents; (2) a resistive layer must be present.

Another crucial feature of the experimental apparatus is the frequency spectrum of the applied e.m. field. Regardless of the reason for which the external e.m. field was originally employed, it seems clear to us that it plays a fundamental role in supplying the energy necessary for the apparent "absorption" of the gravitational field in the disk (compare energetic balance considerations in Section 5). Experimentally one observes [6] that the maximum shielding value is obtained when the coils are supplied with high frequency current (of the order of 10 *MHz*).

Transferring power to the disk is inherently not very efficient. This represents one of the most serious problems to overcome in order to achieve a stable shielding effect, especially for heavy samples. Otherwise, large amounts of refrigerating fluid will be needed to keep the disk from overheating and exiting the superconducting state. Remember that the maximum shielding values have been observed under conditions which seem to trigger the production of heat. This makes clear why a disk made of HTC material is important: although a low-temperature superconductor could also reach the critical density conditions, its specific heat is probably too small to maintain the essential conditions while allowing adequate power transfer.

## 3.2 Interpretation.

We have already stressed in our analysis [4] that an interpretation of the reported effect in the framework of General Relativity, as due to repulsive post-newtonian fields produced by the super-currents [26], is untenable, since the magnitude order of the effect is far too large [27].

Our interpretative model of the experimental results is based on the "anomalous" coupling between Bose condensate and gravitational field described in the previous Sections. In this model the essential ingredient for the shielding is the presence of strong variations of the Cooper pairs density in the disk: *we assume that such variations produce small regions with higher density, where the criticality condition (7) is satisfied*. It is thus necessary to understand, from this point of view, how the experimental parameters summarized above combine to produce such critical conditions.

A correct theoretical interpretation should also suggest alternative experimental settings and possibly lead to a simplified version of the experiment, still giving an observable shielding effect. Some steps towards such a partial replication have been taken by the authors (see below, Section 4, and the Appendix).

We recall that the HTC toroidal disk has a composite structure, with two layers as described in Section 3.1. At the operation temperature (slightly below 70 *K*) the supercurrents, which circulate without any resistance and take the high value $J_c \sim 7000$ $A/cm^2$ in the upper part of the disk, are essentially zero in the lower part. The interface represents a sort of boundary for a fluid motion: the intensity of the supercurrents, which flow parallel to the interface, falls to zero at the interface itself (see Fig. 3).

For our purposes we can regard the supercurrents as a perfect fluid and apply the corresponding equations of fluid dynamics. (Phenomenological theories of superconductivity based on the two-fluids model are largely independent from the detailed microscopic mechanism which produces superconductivity and which is not well known yet in the case of HTCs.) The fluid density - i.e., the density of the Bose condensate - is subjected to local variations, or the fluid is compressible. In other words, while the *average* fraction of superconducting charge carriers in the material depends only on the ratio $T/T_c$ and is small at temperatures about 70 *K*, usually not exceeding a few %, locally there can be for some reason a concentration of superconducting charge carriers, up to a maximum density fixed by the total charge density of the lattice.

Let now *v* represent the local velocity of the superconducting fluid, ρ its local density and *m* the mass of a superconducting charge carrier. If the gradient of *T* can be disregarded and the flow is irrotational (as usually assumed for superfluid motion) the following differential equation holds [28]

grad($v^2/2$ + *V/m*) = - (*kT/m*) grad(logρ/$ρ_0$) ,         (17)

where *k* is the Boltzmann constant and *V* is the external potential, which can be set to zero in the following, since it is almost constant over small distances and in the present case we apply eq. (17) to a very thin region above the interface.

Assuming that ρ and *v* depend only on the coordinate *z*, the solution of (17) is

ρ(*z*) = $ρ_0$ exp[-$mv^2$(*z*)/(2*kT*)].

In order to find ρ as a function of the distance *z* from the interface we need to know the dependence *v*(*z*). Far above the interface, *v* takes an approximately constant value $v_{max}$. We shall assume that *v* vanishes at the interface, due to ohmic resistance, and that *v* grows, close to the interface, as a power of *z*. The behavior of ρ(*z*) is then shown in Fig. 4.

It is seen that the condensate density ρ takes an higher value at the interface than far above it, in the bulk of the "good" superconducting material. This fact is rather intuitive and could have been predicted without use of the fluid equation. Given a "threshold" value $ρ_c$ of the density, there will be a thin layer $0 < z < z_c$ close to the interface in which $ρ_c < ρ(z) < ρ_0$. This layer corresponds to a critical region as defined in Section 1.2. Below the interface (*z* < 0), ρ vanishes. (In practice however, since the interface is not very clear-cut, the curve ρ(*z*) will go smoothly to zero for *z* < 0.)

The relative velocity of the supercurrent with respect to the lower part of the disk is essentially determined by the rotation frequency of the disk [29], because the average velocity of the charge carriers in the supercurrent is on the order of $10^{-5}$ *m/sec*, while the tangential velocity due to rotation is much higher, on the order of 10 *m/sec*. Let us work out this estimate in more detail. The critical current density in the upper layer of the disk is ~ $10^8$ $A/m^2$. Taking a density of charge carriers of $10^{32}$ $m^{-3}$, with an elementary charge of ~ $10^{-19}$ *C* we find *v*=*J*/ρ ~ $10^{-5}$ *m/sec*. On the other hand, when the disk is rotating at angular frequency ω ~ 5000 *rpm* ~ 500 $sec^{-1}$, a point at distance 10 *cm* from the rotation axis has tangential velocity *v*=ω*r* ~ 50 *m/sec*.

We shall briefly discuss in Section 4.1 how conditions similar to those described above, which are present in "stationary form" at the interface in the rotating composite superconducting disk used in [6], are reproduced for short time intervals in our own experiment.

# 4  Our experiment.

One of us (J.S.) has recently succeeded in partially reproducing the weak gravitational shielding effect described in the previous Section. The experimental setup was suggested by the theoretical model outlined in Sections 1-3. It was also designed to eliminate, as far as is possible, any non-gravitational disturbance, and to show a precise temporal correspondence between actions taken on the HTC disk and the weight reduction of the sample/proof mass. Although the observed weight reduction was large enough to be clearly distinguished from the noise (of the order of 1 % or more), it was detected only in transient form, lasting up to ca. 3 *s*. This happened because the weight reduction was coincident with the superconducting transition of the HTC disk, which occurred quite rapidly when the disk warmed up over its critical temperature.

Our earlier experimental setup is described in the Appendix. An improved setup allowed us to run more than 400 trials with an heavier proof mass (a glass rods bundle - total weight 63 *g*), accumulating better statistics. Approximately 10 % of the trials gave positive result, i.e., a clear transient reduction in the weight of the proof mass. This apparently random behavior signals that some of the conditions which are necessary to trigger the effect are not well under control yet. In particular, we found that the duration of the superconducting transition is very crucial: if the transition is too quick or too slow with respect to certain criteria, no effect is observed. The duration of the transition, in turn, depends strongly on the thermal conditions of the disk, and the latter can be controlled only with limited accuracy. A closer discussion of this point can be found in Section 4.1.

Our improved setup did not comprise the AC generator anymore. Thus error analysis is further simplified. One only needs to consider the parasitic effects of buoyancy and molecular diamagnetism. For both error sources, upper limits can be set according to the discussion in [6] and these limits are much smaller (by a factor 20 at least) than the observed effect.

As mentioned above, the transient reduction in the weight of the proof mass was always coincident with the thermal transition of the YBCO disk from the superconducting to the non-superconducting state. This was checked as follows.

A dewar flask with an inside diameter of about 3 *cm* and 14 *cm* deep allows to observe the experiment from the side by virtue of the fact that about 2.5 *cm* of the silvering, inside and out, has been removed when manufactured (see Fig. 5). The top of a hollow cylinder of polyethylene is flush with the bottom of the removed silvering. This cylinder supports, through two parallel bamboo sticks, a 2.5 *cm* by 1.3 *cm*, 0.3 *cm* thick samarium cobalt permanent magnet with an *MGOe* factor of approximately 18. The magnet is even with the bottom of the removed silvering.

The dewar contains liquid nitrogen, as to half fill the clear observation area. This leaves about 1.3 *cm* of empty dewar with the permanent magnet at the bottom. After filling, all is allowed to cool down. A LEVHEX, almost-single-crystal, "pinning type" YBCO hexagon (2.0 *cm* from side to side) is tied to a cotton string and lowered into a second dewar to cool. When the LEVHEX is chilled, it is lowered into the first dewar. The levitation effect is pronounced and a piece of bamboo is used to push the hexagon down and leave it at approximately 0.6 *cm* above the magnet, as opposed to ca. 2.0 *cm* of its highest equilibrium position. The hexagon is well pinned and the familiar "tied to little springs" effect is well in evidence.

The detection system (see Fig. 6) is quite similar in principle to that described in the Appendix. Namely, a 33 *cm* by 51 *cm* glass plate on ring stand clamps is interposed between the dewar and a 63 *g* proof mass made of a hexagonal bundle of 7 rods, each 10 *cm* long and

0.6 *cm* in diameter. The string is tied to a dry 1 *cm* by 1 *cm*, 200 *cm* long wooden beam. The beam acts as a balance and has a 3 *mm* hole drilled in the middle. A 1 *mm* stainless steel pin in an aluminium stirrup serves as a pivot in the middle. The purpose of the beam is to place a "Mettler 300" scale (300 *g* full scale, 0.01 *g* resolution) at least 180 *cm* from the dewar. The proof mass is directly over the levitating hexagon and the far end of the beam has an excess weight (aluminium blocks tied by cotton string) which rests on the scale. Once the whole is set up the scale is tared to leave a weight of 44.0 *g*.

As the liquid nitrogen boils away the temperature of the LEVHEX rises. We see this evidenced by a slight reduction in the height of levitation at first, and, finally, there is no levitation at all. Exactly during this phase, which may last typically between 5 and 10 *s* in the different runs, we observe a transient reduction in the weight of the proof mass as indicated by the counterweight on the balance. In a "positive" run the measurement on the scale generally goes from 44.0 to 46-47 *g* and then returns to baseline after 2-3 *s*. This corresponds to a reduction in the apparent weight of the proof mass [30].

While in Podkletnov's experiment [6] the cryostat and the superconducting disk were enclosed in a stainless-steel box, our apparatus was open, with large aluminum and brass screens placed between the hanging proof masses and the superconducting disk. Although the size of the screens (approx. 1 *m*) was much larger than the mass-disk distance (less than 10 *cm*), it cannot be excluded that transient electromagnetic fields could propagate around the edges of the screen. Nevertheless, we can reasonably exclude that the observed weight variations were due to electromagnetic effects, because:

(1) The proof masses were made of plexiglas, and any metal parts were carefully avoided.

(2) The weight variations were only observed at the superconductive transition. This "timing" rules out electrostatic charge accumulation.

(3) The weight variations were only observed in about 10 % of the trials, while the field of the magnet was always present and changed rapidly during every trial (especially during the transition).

In conclusion, the layout of our experiment is conceived in such a way that it is impossible to enclose the disk into a closed cryostat; on the other hand, it allows to observe a precise temporal coincidence between the state of the disk and the occurrence of gravitational modification.

### 4.1 Interpretation.

A mixed state with locally-enhanced condensate density, similar to that present at the interface of the composite rotating disk (Section 3.2) forms, for a short time interval (a few seconds), also when a good-quality, single-phase HTC superconducting disk warms up above its critical temperature and goes from the superconducting to the normal state.

It is known that at the transition to the normal state some small non-superconducting regions first appear in the superconducting material, and then they gradually grow. Now, if strong circulating supercurrents had been previously induced in the material, these currents will rapidly decay at the transition. However, there will be during the transition a short intermediate phase in which part of the supercurrents still circulate, through those regions in which the material is still superconducting (Fig. 7). During this phase the flow of the supercurrents can still be described by a perfect fluid model like in Section 3.2. The non-superconducting regions act as boundaries for the flow and the resulting pattern for the condensate density will exhibit roughly similar features to those we found earlier: namely, the condensate density will be higher near the resistive obstacles, where the velocity must be smaller. Clearly if the superconducting transition is "too slow", and the system is allowed to reach equilibrium, the critical density may be not achieved anywhere.

Unlike in the case of the rotating toroid, the whole bulk of the HTC disk is involved in this case. It is crucial for the occurrence of the effect that the supercurrent present in the disk at the transition be very high and that the transition width $\Delta T_c$ be small. This requires an excellent HTC ceramic material.

The application of AC magnetic fields may allow some control on the transition rate. We are presently improving our earlier apparatus (see the Appendix) in this direction. The method described above, however, can be used to demonstrate the effect in a very clean way.

## 5 Energetic balance.

It is necessary now to discuss some issues of elementary character - but important from the practical point of view - concerning the overall energetic balance of the shielding process.

In general energy must be supplied in order to reduce the weight of an object, because the potential gravitational energy of the object has negative sign and is smaller, in absolute value, in the presence of shielding. However, since the field is non-conservative, it is certainly wrong to compute the difference in the potential energy of an object between the interior and the exterior of the shielding cylinder by evaluating naively the difference (which turns out to be huge) between an hypothetical "internal potential"

$$U = - GMM_E(1-\alpha)/R_E = - Mg_E R_E(1-\alpha),$$

and an "external potential" $U=-Mg_E R_E$. (Here $M$ is the mass of the object, $R_E$ the Earth radius, $M_E$ the Earth mass and $\alpha$ the "shielding factor", $\alpha \sim 0.01$.)

Moreover, the gravitational fields with which we are most familiar, being produced by very large masses, are relatively insensitive to the presence of light test bodies; thus, it makes sense in that case to speak of a field in the usual meaning: while a body falls down, we do not usually worry about its reaction on the Earth. But in the present case, the interaction between the shielded object and the external source (that is, the system [Bose condensate+external e.m. field], which, by fixing the constraints on the gravitational potential $h$, produces the shielding), is very important [31].

Let us then ask a provocative question, suggested by the experimental reality: if the superconducting disk is in a room and the shielding effect extends up to the ceiling, should we expect the disk and all the shielding apparatus to feel a back reaction? (And possibly an even stronger one if the ceiling is quite thick or if there are more floors above?) The most reasonable answer is this: since the ceiling is very rigid, the experimental apparatus is not able to exert any work on it and thus does not feel its presence.

To clarify this point, imagine that we hang over the superconducting disk a spring of elastic constant $k$ and holding a body of mass $M$ at rest. Then we operate on the disk with proper e.m. fields and produce the shielding effect with factor $\alpha$. Now, the gravity acceleration over the disk becomes $g_E(1-\alpha)$. If the shielding effect is obtained quickly, the mass should begin to oscillate, otherwise it should rise by an height of $\Delta x = \alpha g_E M k^{-1}$, while remaining in equilibrium. In any case, since the kinetic energy and the potential energy of a harmonic oscillator in motion have the same mean value, it is legitimate to conclude that the work done by the shielding apparatus on the system [mass+spring] is on the order of

$$\Delta E \sim k\, (\Delta x)^2 \sim (\alpha g_E M)^2\, k^{-1}. \qquad (19)$$

This example shows that the work exerted by the apparatus on a sample being "shielded" will depend in general on the response of the sample itself. Only if the object being shielded can respond to the effect will work be done.

At this point we can estimate how much energy is needed to achieve critical density in one region of the condensate of cross section $\sigma$. If $\sigma_{sample}$ is the projection of the sample on the disk, this energy is given by

$$\Delta E_\sigma = (\sigma/\sigma_{sample})\Delta E \sim (\sigma/\sigma_{sample})(\alpha g_E M)^2\, k^{-1}.$$

This energy must be supplied by the external variable e.m. field.

In conclusion, we must expect in general an interaction between the partially shielded samples and the shielding apparatus. The energy needed to shield a sample depends on the mass of the sample itself and on the way it is constrained to move. In particular, we deduce from eq. (19) that if we want to detect the shielding effect by measuring the deformation of a spring (and do this with the minimum influence on the shielding apparatus) we should use, as far as is allowed by the sensitivity of the transducer, a spring with a high rigidity coefficient $k$.

**Acknowledgments.**

We are very grateful to E. Podkletnov for advice and encouragement. We are very much indebted to J.R. Gaines, Vice President of Superconductive Components, Columbus, Ohio, USA, who supplied the HTC disk.

Sitter background [11]. In our case, however, we consider only localized modifications to the lagrangian. Thus neither objection is relevant.

18. Short-distance regulators or higher-order terms, analogous to a "$\Phi^4$" term in the scalar field case, are present in several models of quantum gravity [10,12]. Note that the size of our hypothesized critical regions can be regarded as macroscopic with respect to the scale where conformal fluctuations of the metric or $R^2$-terms become important. For this reason, we believe that an approach based upon the Euclidean weak field approximation plus regulator is appropriate.

19. G. Modanese, J. Math. Phys. **40** (1999) 3300.

20. G. Modanese, *Gravitational Anomalies by HTC Superconductors: a 1999 Theoretical Status Report*, report physics/9901011, January 1999.

21. G. Modanese, *Phys. Lett.* **B 460** (1999) 276; *Nucl. Phys.* **B 588** (2000) 419; *Phys. Rev.* **D 62** (2000) 087502.

22. I.J. Muzinich, S. Vokos, Phys. Rev. **D 52** (1995) 3472.

23. Ning Li, D. Noever, T. Robertson, R. Koczor and W. Brantley, Physica **C 281** (1997) 260; R. Koczor and D. Noever, *Fabrication of Large Bulk Ceramic Superconductor Disks for Gravity Modification Experiments and Performance of YBCO Disks Under e.m. Field Excitation*, Report AIAA 99-2147.

24. C.S. Unnikrishnan, Physica **C 266** (1996) 133.

25. G. Modanese, *Updating the theoretical analysis of the weak gravitational shielding experiment*, report UTF-367/96.

26. D.G. Torr and Ning Li, Found. Phys. Lett. **6** (1993) 371.

27. The work [24] shows that the contribution from the super-currents to the static component $g_{00}$ of the post-newtonian gravitational field over the disk is not only much smaller than the observed effect (by several magnitude orders), but it is attractive like the newtonian field of the Earth. Even taking into account perturbative quantum corrections to the Newton potential one reaches the same negative conclusion [4].

28. K. Huang, *Statistical Mechanics*, Ch. 5 (J. Wiley, New York, 1963).

29. We assume that the superconducting charge carriers are not able to follow the rotation of the disk, or only partially: those very close to the interface co-rotate with the disk, those farther away follow slower, with a relative velocity $v_{max}$. This behavior is also confirmed by the observations relating to the "braking phase", during which the rotation frequency of the disk is suddenly reduced [6].

30. The apparent shielding factor can depend not only on the intrinsic weight change, but also on the frequency response of the detecting apparatus. Namely, if the transient weight variation occurs as a series of short spikes, it can resonate with the detector. This might explain why a recent replication of our experiment, with a different detection method, gave clear but smaller values for the transient weight change

## Appendix: earlier set-up.

Our first experimental set-up consisted of (a) a hexagon-shaped YBCO 1" (2.5 *cm*) superconducting disk, 6 *mm* thick; (b) a magnetic field generator producing a 600 *gauss*/60 *Hz* e.m. field; (c) a beam balance with suspended sample.

The beam is made of bamboo, without any metal part, coming to a point on one end (24.6 *cm* long, weight 1.865 *g*). The sample is made as follows. A cardboard rectangle (16 *mm* by 10 *mm* by 0.13 *mm*) is suspended from the balance with 2.8 *cm* of cotton string. A polystyrene "pan" (7.2 *cm* by 8.7 *cm* by 1.7 *mm*) is attached with paper masking tape to the cardboard rectangle. The total sample assembly (with string, cardboard, tape) weighs 1.650 *g*.

The balance is suspended from the end of a 150 *cm* wood crossbeam by ca. 30 *cm* of monofilament fishing line attached to the balance's center of mass (5.5 *cm* from the end where the sample is attached). The other end of the crossbeam is firmly anchored by a heavy steel tripod. Thermal and e.m. isolation is provided by a glass plate (15 *cm* by 30 *cm*, 0.7 *cm* thick) with a brass screen attachment. This plate-and-brass-screen assembly is held about 4.5 *cm* below the sample by a "3-finger" ring stand clamp. A straightsided, 14 *cm* diameter, 25 *cm* deep dewar with ca. 10 *cm* of liquid nitrogen is used to cool the superconducting disk below its critical temperature, and is removed from the experiment area before the trial.

The experimental procedure comprises the following steps.

1. The YBCO superconductor is placed in a liquid nitrogen bath and allowed to come to liquid nitrogen temperature (as indicated when the boiling of the liquid nitrogen ceases). The superconductor will remain below its critical temperature (about 90 *K*) for the duration of the trial (less than 20 seconds).
2. The disk is then removed from the bath and placed on a strong NdFeB magnet to induce a supercurrent. The Meissner effect is counteracted by a wooden stick. The superconducting disk has a cotton string attached to it to assist handling.
3. The disk and wooden stick assembly is placed on the AC field generator, about 33 *cm* below the isolation plate and about 40 *cm* below the sample. The AC field generator is then cycled for ca. 5 seconds with 0.75 *sec* equal-time on/off pulses. Prior to a run the sample is centered to be over the middle of where the disk will finally be, on the AC field generator.

One observes that while AC current is flowing through the generator the balance pointer dips 2.1 *mm* downward. When the AC field generator is pulsed with no superconductor present, there is no measurable pointer deflection. Also air flows do not cause any measurable deflection. The whole procedure is well reproducible.

The weight difference required to raise the sample by 2.1 *mm* was then found to be 0.089 *g*. This was measured taking advantage of the fact that the suspension wire produces a small torque on the balance beam toward the equilibrium position: the balance pointer was found to raise 2.1 *mm* upward when a weight of 0.089 *g* was placed above the sample.

# FIGURE CAPTIONS

1. *Typical double-well potential of a ($\Phi^4 - \Phi^2$) field system.*

   The curve A represents a familiar "$\Phi^2$" potential, corresponding to a field with positive squared mass term. The curve B represents the same "$\Phi^2$" potential, to which a "$\Phi^4$" term has been added. The lowest energy excitations are nearly the same for A and B. In C, the sign of the "$\Phi^2$" term has been reversed (($\Phi^4 - \Phi^2$) potential). Now, in the lowest excitations the field is pinned to finite values distinct from zero.

2. *Non-conservative character of the "modified" field above the superconducting disk.*

   The "modified field" described by Podkletnov, with an infinite cylinder-like shielding zone over the disk, is clearly non-conservative. This means that if a proof mass goes along a closed path (up within the shielding zone, and down outside), the gravitational field exerts a net positive work on it.

   From the formal point of view, this means that the observed static field is not the gradient of a gravitational potential, even though the Einstein equation for the connection (div$\Gamma$=0) still holds true.

   Physically, this property is due to the ("anomalous") coupling of the gravitational field with the external e.m. field, coupling that is mediated by the Bose condensate in the superconducting disk. The mechanical energy that a proof mass gains when it goes along the closed circuit is furnished by the external e.m. field.

3. *Schematic diagram representing the velocity distribution of the superconducting charge carriers in the upper layer of the toroidal disk (B) during rotation.*

   The lower layer (A) is not superconducting at the temperature of operation (slightly below 70 *K*). The *relative* velocity of the superconducting charge carriers with respect to the interface between the two layers is represented by thick arrows. It grows from zero to some velocity $v_{max}$ according to a power law (compare Fig. 4).

   This is a side view. In stationary conditions, the relative velocity has no component along the vertical direction *z*. We suppose that the disk is rotating clockwise, so the short arrow means in reality that the superconducting charge carriers near the interface are almost co-rotating with the disk; the carriers farther away from the interface "follow slower", thus with larger relative velocity, up to $v_{max}$.

4. *Variation of the condensate density in the upper layer of the disk as a function of the distance from the interface, according to the perfect-fluid model.*

For the definition of the coordinate $z$ and a qualitative description of the velocity pattern compare Fig. 3.

The curve $v(z)/v_{max}$ shows for illustration purposes a quasi-linear growth of the velocity of the superconducting charge carriers with increasing distance from the interface, followed by saturation to the value $v_{max}$. The curves A, B, C represent the behavior of the local condensate density $2\rho/\rho_0$ (the factor 2 is for graphical reasons) according to eq. (18), for the following specific cases:

A: $v(z)/v_{max} \sim z^{1/2}$

B: $v(z)/v_{max} \sim z$

C: $v(z)/v_{max} \sim z^{3/2}$

It is seen that the condensate density achieves its maximum value for $v=0$ and its minimum for $v=v_{max}$.

The line D represents the threshold condensate density $\rho_c$ necessary for anomalous coupling. $Z_c$ represents the thickness of the critical region of the condensate if the velocity depends on $z$ like in A (the generalization to the cases B and C is obvious). This means that in the thin layer $0 < z < z_c$ close to the interface one has $\rho_c < \rho(z) < \rho_0$, or the layer is a "critical region". Note that the value of the ratio $\rho_c/\rho_0$ as depicted above and the $z$-scale are just for illustration purposes.

5. *Scheme of our [magnet+LEVHEX] setup.*

    A: Dewar flask (height 14 *cm*, inside diameter 3 *cm*).

    B: Part of the dewar (height 2.5 *cm*) where the silvering has been removed to allow side observation of the levitating YBCO hexagon (E).

    C: Polyetilene tube supporting the magnet (D) through two parallel bamboo sticks.

    D: samarium cobalt permanent magnet (2.5 *cm* by 1.3 *cm*, 0.3 *cm* thick).

    E: LEVHEX "pinning type" YBCO hexagon (2.0 *cm* from side to side).

6. *Detection system for the demonstration experiment of transient shielding effect at the superconducting transition.*

    A: Dewar flask (see details in Fig. 5) with LEVHEX levitating above magnet.

B: Glass screen, with brass and aluminum foils, to shield proof mass from air flows and electrostatic fields.

C: Proof mass: bundle of 7 glass rods, each 10 *cm* long and 0.6 *cm* in diameter; total weight 63 *g*.

D: Wooden balance beam, 1x1x200 *cm*.

E: Mettler 300 balance, 0.01 *g* resolution.

7. *Velocity and density pattern of the superconducting charge carriers in a bulk disk at the transition (dark: critical regions).*

   At the transition to the normal state some small non-superconducting regions first appear in the superconducting material, and then they gradually grow. If strong circulating supercurrents had been previously induced in the material (top), these currents will rapidly decay at the transition. However, there will be during the transition a short intermediate phase in which part of the supercurrents still circulate, through those regions in which the material is still superconducting (bottom).

   During this phase the flow of the supercurrents can still be described by a perfect fluid model like in Section 3.2. The non-superconducting regions act as boundaries for the flow and the resulting pattern of the condensate density will exhibit roughly similar features to those we found earlier: namely, the condensate density will be higher near the resistive obstacles, where the velocity must be smaller.

Figure 1

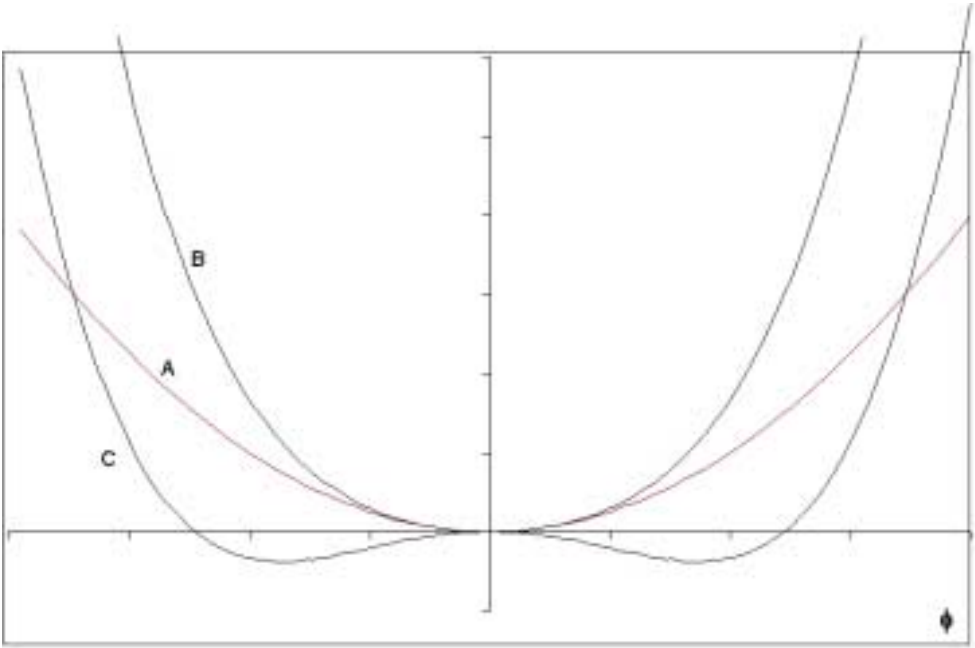

Figure 2

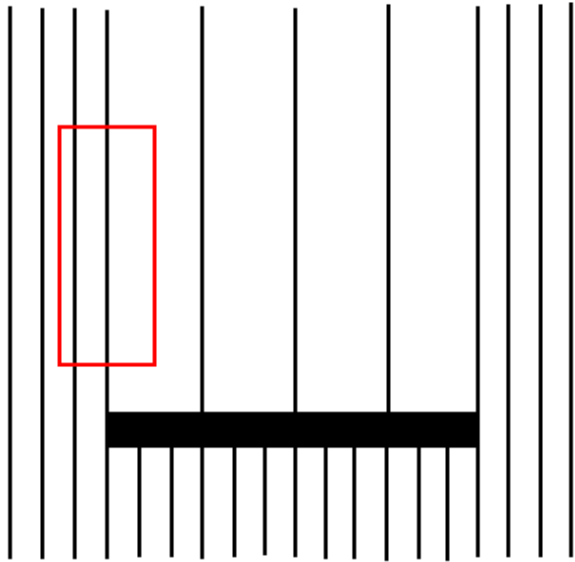

Figure 3

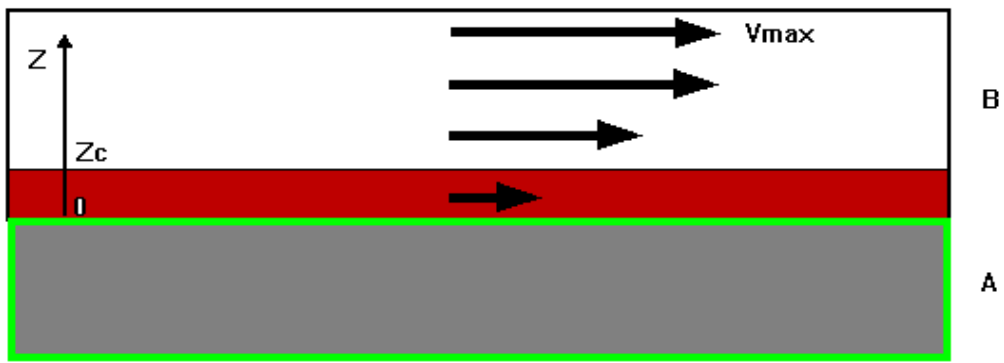

Figure 4

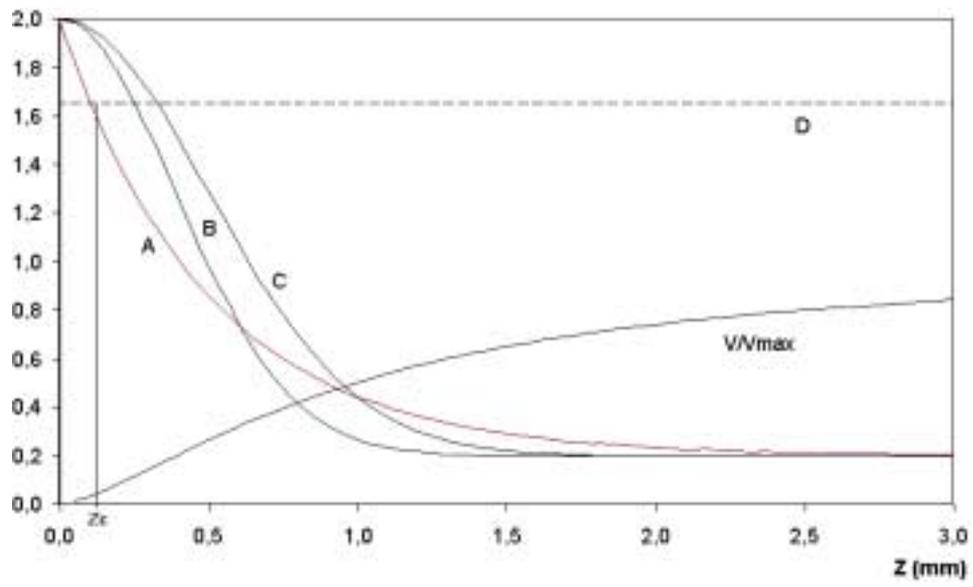

Figure 5

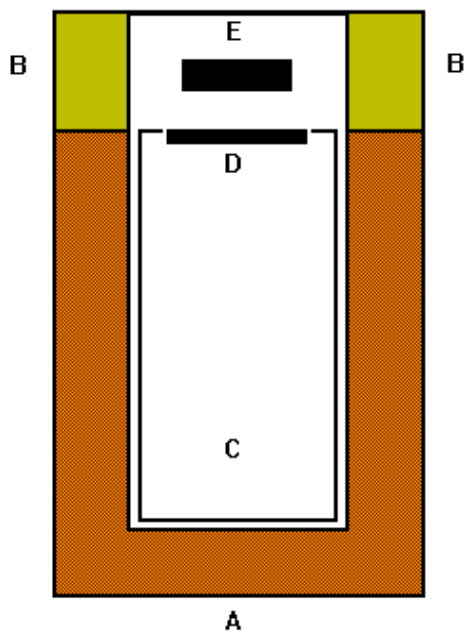

Figure 6

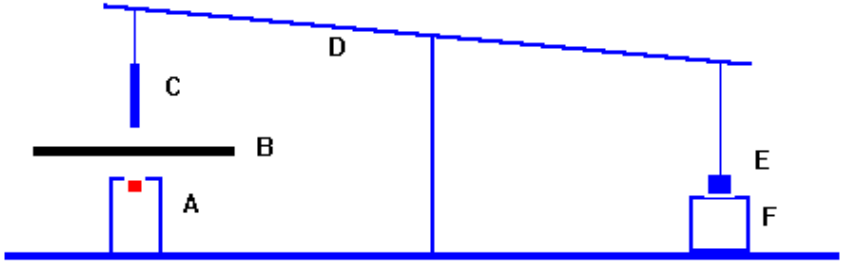

Figure 7

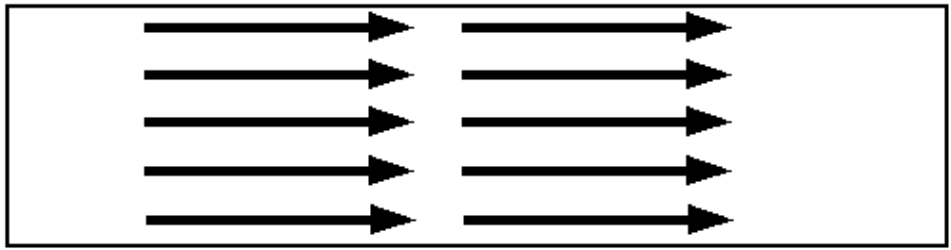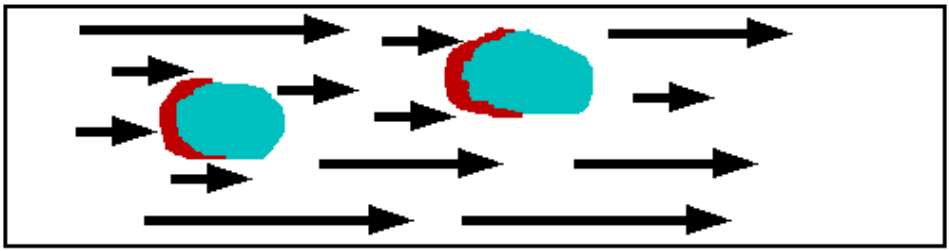